\documentclass[acmtog]{acmart}
\acmSubmissionID{521}

\usepackage{graphicx} 
\usepackage{booktabs} 
\usepackage{soul}                    
\usepackage{multirow}

\def\fvvdp{Fov\-Video\-VDP}
\def\stela{StelaCSF}
\def\csf{S}
\def\fs{f_{s}}
\def\load{a}
\def\ecc{e}
\def\gain{g}
\def\thr{t}

\def\ahigh{\textrm{``high''}$\,$}
\def\amed{\textrm{``medium''}$\,$}
\def\alow{\textrm{``low''}$\,$}

\graphicspath{{figures/}}

\usepackage[ruled]{algorithm2e} 

\SetAlFnt{\small}
\SetAlCapFnt{\small}
\SetAlCapNameFnt{\small}
\SetAlCapHSkip{0pt}

\setcopyright{none}
\renewcommand\footnotetextcopyrightpermission[1]{}
\setcopyright{acmlicensed}
\acmJournal{TOG}
\acmYear{2023}
\acmVolume{42}
\acmNumber{4}
\acmArticle{1}
\acmMonth{8}
\acmPrice{15.00}
\acmDOI{10.1145/3592406}

\begin{document}
\title{Towards Attention-aware Foveated Rendering}

\author{Brooke Krajancich}
\affiliation{%
  \institution{Stanford University}
   \country{USA}
}
\email{brookek@stanford.edu}
\author{Petr Kellnhofer}
\affiliation{%
  \institution{TU Delft}  
   \country{Netherlands}
}
\email{p.kellnhofer@tudelft.nl}
\author{Gordon Wetzstein}
\affiliation{%
 \institution{Stanford University}
  \country{USA}
}
\email{gordon.wetzstein@stanford.edu}

\renewcommand\shortauthors{Krajancich, et al.}

\begin{abstract}
Foveated graphics is a promising approach to solving the bandwidth challenges of immersive virtual and augmented reality displays by exploiting the falloff in spatial acuity in the periphery of the visual field.
However, the perceptual models used in these applications neglect the effects of higher-level cognitive processing, namely the allocation of visual attention, and are thus overestimating sensitivity in the periphery in many scenarios.
Here, we introduce the first attention-aware model of contrast sensitivity.
We conduct user studies to measure contrast sensitivity under different attention distributions and show that sensitivity in the periphery drops significantly when the user is required to allocate attention to the fovea.
We motivate the development of future foveation models with another user study and demonstrate that tolerance for foveation in the periphery is significantly higher when the user is concentrating on a task in the fovea. 
Analysis of our model predicts significant bandwidth savings over those afforded by current models.
As such, our work forms the foundation for attention-aware foveated graphics techniques.
\end{abstract}

%
%


%
%


  \begin{teaserfigure}
   \includegraphics[width=\textwidth]{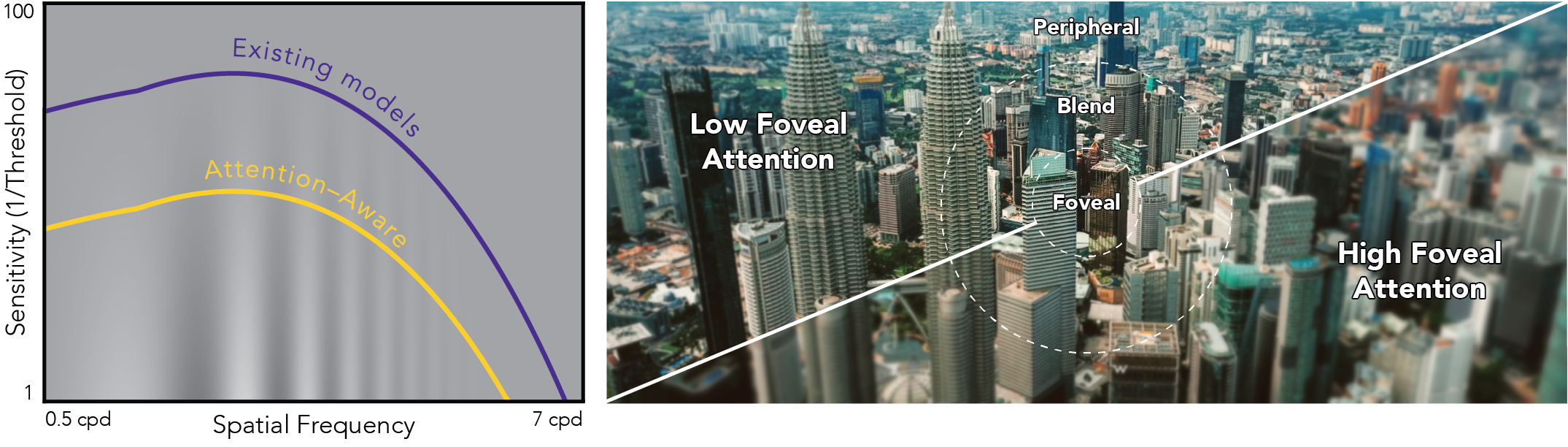}
    \caption{Foveated graphics techniques rely on eccentricity-dependent models of human vision. However, existing models of contrast sensitivity (left, purple line, shown for a fixed eccentricity) do not take into account allocation of visual attention across the visual field. Our work is the first to experimentally derive a model for eccentricity-dependent attention-aware sensitivity (left, yellow line). As illustrated on the right, when the user is focused on a task in the fovea, less attention is directed to the periphery and a higher level of foveation (i.e., peripheral blur) is possible without impacting the perceived visual quality.}
  \end{teaserfigure}

\maketitle
\fancyfoot{}
\thispagestyle{empty}

\section{Introduction}
\label{sec:intro}
Virtual and augmented reality (VR/AR) are next-generation display systems that promise perceptually realistic user experiences by matching the resolution of human vision across a wide field of view.
However, the necessary bandwidth for rendering, streaming, and displaying the required visual data is not yet possible with current technology.
Foveated graphics has emerged as a suite of techniques that exploits the eccentricity dependent acuity of human vision to minimize bandwidth in an imperceptible manner.
In VR/AR, this is often implemented using gaze-contingent rendering, shading, compression or display (see Sec.~\ref{sec:lit_foveated}).
While these methods build on the insight that the human visual system has a limited ability to sense spatio-temporal changes in light, they have yet to consider how this might be dependent on higher-level cognitive processing.

Indeed, research shows that we rarely see what we are looking at unless we direct sufficient cognitive resources~\cite{mack2003inattentional}. This explains many phenomena, including change blindness or tunnel vision.
\emph{Visual attention} refers to a set of cognitive operations that helps us selectively process the vast amounts of information with which we are confronted, allowing us to focus on a certain location or aspect of the visual scene, while ignoring others.
Most often, we direct our attention \emph{overtly}, by moving our eyes towards a location, but we can also direct attention to an area in the periphery \emph{covertly}, via a mental shift.
Several studies have demonstrated that, under many conditions, increasing the amount of attention allocated to a visual task can enhance performance ~\cite{lee1997spatial, sperling1978attention}.
In a similar manner, dividing attention between tasks reduces contrast sensitivity~\cite{mahjoob2019contrast, huang2005attentional}, visual acuity~\cite{mahjoob2022effect}, and speed of information accrual~\cite{carrasco2006attention}.

However, existing models of contrast sensitivity and visual acuity across the visual field are built on experiments where subjects are asked to covertly direct high levels of visual attention to a discrimination task in the periphery.
Thus, for most scenarios in the real world and VR/AR, where most of our attention is directed overtly (at our gaze position), we are likely overestimating our perceptual abilities in the periphery.
Consequently, current efficacies of foveated graphics are too conservative in most real-use cases.

In this paper, we propose to account for the effect of covert attention when modeling human contrast sensitivity.
To this goal, we investigate the effect of modulating the amount of attention allocated to the contrast discrimination task in the periphery, by forcing attention to the fovea with a visually demanding task.
Specifically, we compare the standard approach to measuring contrast sensitivity, where a low amount of attention is directed to the fovea (\alow), to scenarios where part or most of the attention is directed there (\amed and \ahigh).
We show that in such instances, peripheral contrast discrimination thresholds elevate significantly, introduce the first attention-aware contrast sensitivity model, and motivate the development of future foveation models that take this into account.

To summarize, we make the following contributions:
\begin{itemize}	
\item We design and conduct user studies to measure and validate eccentricity-dependent effects of attention on contrast discrimination and foveation efficacy.
\item We introduce the first analytic model of contrast sensitivity across eccentricity under varying attention.
\item We analyze bandwidth considerations and demonstrate that our model may afford significant bandwidth savings over existing fo\-ve\-a\-ted graphics techniques.
\end{itemize}

\paragraph{Overview of Limitations}
The primary goals of this work are to develop the first perceptual model for attention-dependent contrast sensitivity and to demonstrate its potential benefits to foveated graphics. However, we do not attempt to derive a measurement instrument for attention allocation across the visual field, nor do we propose new foveated rendering algorithms or specific compression schemes that directly use this model.

\section{Related Work}
\label{sec:related}
\subsection{Foveated Graphics}
\label{sec:lit_foveated}

Foveated graphics techniques exploit eccentricity-dependent aspects of human vision, such as acuity, to minimize the bandwidth of a graphics system by optimizing bit depth~\cite{mccarthy_sharp_2004}, color-fidelity~\cite{duinkharjav2022color}, level-of-detail~\cite{luebke_perceptually_2001, murphy_gaze-contingent_2001}, or by simply reducing the number of vertices or fragments a graphics processing unit has to sample, ray-trace, shade, or transmit to the display; see~\cite{Duchowski:2004,Koulieris:2019} for a review of this area. Foveated rendering is perhaps the most well-known example of this class of algorithms~\cite{geisler_real-time_1998,guenter_foveated_2012,patney_towards_2016,sun_perceptually-guided_2017,friston2019perceptual,deng2022fov,Kaplanyan_2019_DeepFovea,tursun2019luminance,tariq2022noise}. The perceptual models underlying foveated graphics usually exploit spatial aspects of eccentricity-dependent human vision but, to the best of our knowledge, none of them model cognitive or attentional effects of human vision, which we aim to address with our work.

\subsection{Eccentricity-dependent CSF Models}
\label{sec:lit_csf}
The human visual system (HVS) is limited in its ability to sense variations in light intensity over space and time. 
Such visual performance is often described by the spatio-temporal contrast sensitivity function (CSF), defined as the inverse of the contrast discrimination threshold, that is, the smallest contrast of sinusoidal grating that can be perceived at each spatial and temporal frequency~\cite{robson1966spatial}. 
The CSF can be used to describe the gamut of visible spatio-temporal signals as well as the decrease in relative sensitivity with retinal eccentricity.

While the CSF has been studied for over 70 years~\cite{robson1966spatial}, Kelly~\shortcite{kelly_motion_1979} was the first to fit an analytical function, although limited to the fovea and a single luminance. 
Similarly, Watson et al.~\shortcite{watson2016pyramid} devised the pyramid of visibility, a simplified model that can be used if only higher frequencies are relevant. 
This model also captured luminance dependence and was later refit to model stationary content at higher eccentricities~\cite{watson2018field}. 
Recently, models capturing eccentricity dependence for the full spatio-temporal domain were also presented~\cite{krajancich2021perceptual,tursun2022perceptual,mantiuk2021fovvideovdp}. 
Notably, Mantiuk~et~al.~\shortcite{mantiuk2022stelaCSF} proposed a unified model, \stela{}, which accounts for all major dimensions of the stimulus: spatial and temporal frequency, eccentricity, luminance, and area by combining data from several previous papers.
Similar to luminance contrast, sensitivity to color contrast has also been studied and a spatio-chromatic CSF for foveal vision has recently been fitted~\cite{Mantiuk2020PracticalCC}.

However, the data used to fit eccentricity-dependent models is collected from experiments where subjects covertly direct high levels of visual attention to the contrast discrimination task in the periphery. Such a scenario is unlikely to be representative of viewing conditions in the real world or in VR/AR settings, and may overestimate perceptual thresholds.

A related perceptual quality relevant for foveation is visual acuity defined as the smallest resolvable image detail.
While our work focuses on measurements of the CSF alone, we expect similar effects to apply to acuity as well, because acuity can be understood as the outer limit of the CSF gamut where the sensitivity of vision drops sharply.

\subsection{Attention--dependent CSF}
\label{sec:lit_attention}

Visual attention lies at the crossroads between perception and cognition, allowing us to select relevant sensory information for preferential processing.
It is often modeled as a ``zoom'' or "variable-power lens'', that is, the attended region can be adjusted in size, but defines a tradeoff between its size and processing efficiency because of limited processing capacities~\cite{eriksen1986visual}.
Physiologically, attention modulates neuronal responses and alters the profile and position of receptive fields near the attended location~\cite{anton2013attentional}.
Behaviorally, it improves performance in various visual tasks. One prominent effect of attention is the modulation of performance in tasks that involve the visual system's spatial resolving capacity~\cite{carrasco2018visual}.

In line with the ``zoom lens'' model, several studies have shown that covert attention enhances contrast sensitivity at the attended location at the cost of decreased sensitivity at unattended locations across the visual field, at different eccentricities and isoeccentric (polar angle) locations~\cite{carrasco2000spatial, cameron2002covert, mahjoob2019contrast, carrasco2011visual}. In such studies, attention is typically modulated by visual means, e.g., pre-cueing the location of the visual target~\cite{carrasco2000spatial, cameron2002covert}, or by drawing attention away from the stimuli by use of a concurrent visual task presented elsewhere~\shortcite{huang2005attentional, morrone2004different}. Carrasco~et~al.~\shortcite{carrasco2000spatial} found that pre-cuing attention to the visual target enhanced contrast sensitivity between ~0.05 and 0.1 log units over a broad range of spatial frequencies, and later, Carrasco~et~al.~\shortcite{ling2006sustained} described this attention effect as equivalent to applying an effective \emph{contrast gain} to the stimulus. A similar effect also occurs for visual acuity~\cite{montagna2009attention} and speed of information accrual~\cite{carrasco2006attention}. 

Also in line with the ``zoom lens'' model, and most similar to our work, Huang and Dobkins~\shortcite{huang2005attentional} showed that when attention is divided across several points in the visual field, this reduces its enhancement effect at each location. In particular, they showed that drawing attention to the fovea with a rapid serial visual presentation task reduced contrast discrimination performance in the periphery by up to a factor of 10. 

However, each of these studies measures a single position in eccentricity in the perifovea ($5-10^\circ$) and often only for 1 or 2 users. To the best of our knowledge, this effect has not been modeled over the visual field, nor is there any available data for the effect in the periphery ($>10^\circ$), which we hope to rectify with our work.

\section{A Model for Perception Under Divided Attention}
\label{sec:study}
While modulating the amount of attention has been shown to affect contrast discrimination thresholds (see Sec.~\ref{sec:lit_attention}), insufficient data and a lack of existing models prevent this effect from being applied to existing CSF models and hence foveated graphics. In this section, we provide a detailed discussion of the user study we conducted and the model we fit to predict the effect of modulating peripheral attention on the CSF.

\subsection{Measuring CSF}
\label{sec:csf_study}

The CSF model we wish to acquire could be parameterized by temporal frequency, spatial frequency, rotation angle, eccentricity (i.e., distance from the fovea), direction from the fovea (i.e., temporal, nasal, etc.) and other parameters. 
However, due to the fact that each data point needs to be recorded for multiple subjects and for many contrasts per subject to determine the respective CSFs, sampling all dimensions at once seems infeasible. Instead, we nominally select 3 points across the eccentricity ($\ecc$) available with our display (see Sec.~\ref{sec:model_user_study}), a spatial frequency ($\fs$) of 2~cpd, and a diameter of $5^\circ$ for the furthest point. We then use the \emph{cortical magnification factor} to scale the spatial frequencies and diameters at the other retinal positions (see stimuli No. 1--3 in Table~\ref{tab:stimuli}) such that the discrimination thresholds should be approximately the same (see Supplement for more detail).

In order to obtain the contrast discrimination thresholds, we use Gabor patches, i.e., sinusoidal gratings modulated by a Gaussian function (see Supplement for more detail), as is used by most previous works (Table 1, \cite{mantiuk2022stelaCSF}). 
During each trial, 2 patches are simultaneously presented for 500\,ms, centered at a given eccentricity, to the left and right side of the central fixation position. Each grating is randomly orientated either horizontally or vertically and the user is asked to discriminate whether the patches are of the same or different orientations.

\begin{table}
\caption{\label{tab:stimuli}Parameters of tested Gabor patches. For measuring the model (shown above the divider), we chose a diameter of $5^\circ$ at the highest eccentricity of $21^\circ$ to utilize the full field of view of our display and a spatial frequency $\fs$ of 2~cpd, then use the cortical magnification factor to scale these parameters at $7^\circ$ and $14^\circ$ eccentricity. Gabor's sigma was defined as 20\% of the diameter. For validation, we chose 2 sets of Gabor parameters (shown below the divider) used to fit StelaCSF~\shortcite{mantiuk2022stelaCSF}, namely measurements taken by Virsu and Rovamo \shortcite{virsu1979visual} and Wright~and~Johnson~\shortcite{wright1983spatiotemporal}. Stimulus No. 5 was also tested at 2 additional adaption luminances, $58$ and $116$\,cd/m$^2$ (No. 6 and 7).
}
\begin{center}
\begin{tabular}{ccccc} 
\toprule
  No. & 
  \begin{tabular}[c]{@{}c@{}}Eccentricity \\ ($^\circ$)\end{tabular}  & 
  \begin{tabular}[c]{@{}c@{}}Diameter \\ ($^\circ$)\end{tabular}  & 
  \begin{tabular}[c]{@{}c@{}}Spatial \\ Freq. (cpd)\end{tabular} & 
  \begin{tabular}[c]{@{}c@{}}Adaptation \\ Lum. (cd/m$^2$)\end{tabular}  \\
\midrule
 1 & 7 &  2.16 & 4.62 & 28  \\ 
 2 & 14 &  3.58 & 2.79 & 28  \\ 
 3 & 21 & 5 & 2 & 28 \\ 
 \midrule
 4 & 9.25 & 1.7 & 2 & 28 \\ 
 5 & 15 &  5 & 4 & 28 \\ 
 6 & 15 &  5 & 4 & 58 \\ 
 7 & 15 &  5 & 4 & 116 \\
\bottomrule
\end{tabular}
\end{center}
\end{table}

\subsection{The Attention-modulating Task}
\label{sec:att_mod_task}
Inspired by Huang and Dobkins~\shortcite{huang2005attentional}, we present a rapid serial visual presentation (RSVP) at the fixation cross in order to modulate the amount of attention paid to the peripheral contrast discrimination task. The RSVP stimulus consists of $N$ $1^\circ \times 1^\circ$ letters, each lasting $500/N$\,ms with 0\,ms blank in between, such that the task lasts the total display duration of the peripheral Gabor patches. The color of the letters alternate between red and green (scaled to be approximately isoluminant with the background), where the initial color is randomized across trials, and the user is asked to identify the color of the ``target letter'' (the letter ``T'', which appears only once in a given sequence). Increasing $N$ increases the difficulty of the task and should force more attention to the fovea, at the cost of reduced attention to the periphery.
Consequently, three task levels were chosen to have an $N$ of 1 (easy), 4 (medium) and 6 (hard), to force \alow, \amed, and \ahigh levels of attention to the fovea.
The target letter ``T'' was also adjusted such that for the \amed and \ahigh attention tasks it would not appear in the first 3rd of letters to avoid users obtaining the color early enough to shift their attention to the periphery before the trial ended.

\subsection{User Study}
\label{sec:model_user_study}

\begin{figure}[t!]
   \includegraphics[width=0.8\linewidth]{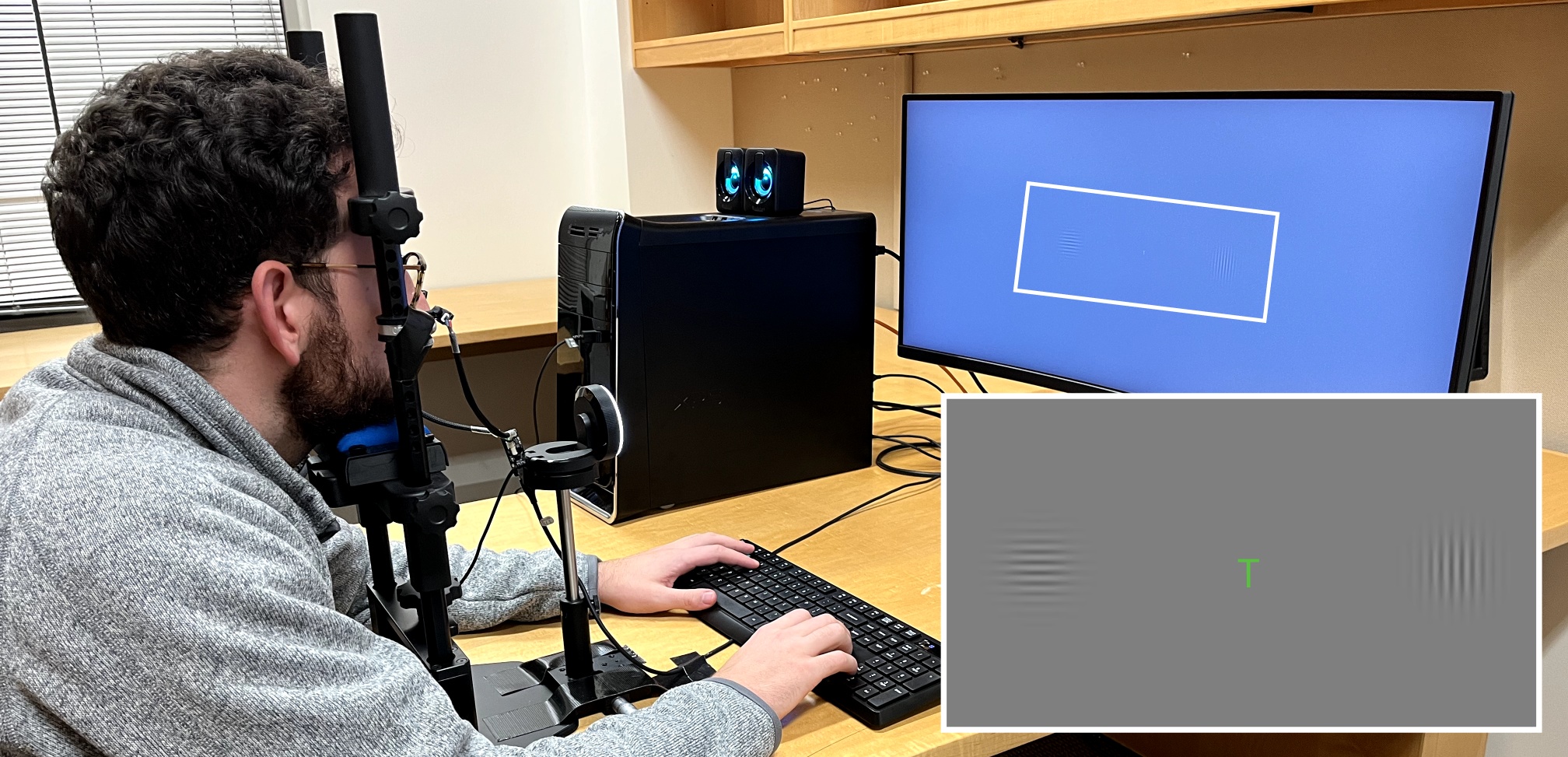}
    \caption{Photograph of the user study setup. The inset shows an enlarged illustration of the stimulus on the screen; the central RSVP letter task with the Gabor patches centered at $e$ to the left and right. The brightness of the letter T has been exaggerated for visibility.}
    \label{fig:setup}
\end{figure}

\paragraph{Setup}
Due to the need to display high resolution stimuli across a wide field of view, we conduct our study using a 34~inch, 144~Hz Dell Curved Gaming Monitor (Model No. S3422DWG, see Fig.~\ref{fig:setup}).
This display has an adjustable backlight, allowing us to tune luminance. For this study we use a setting that gives a minimum and maximum luminance of $0.6$\,cd/m$^2$ and $104$\,cd/m$^2$, respectively, and a gamma of 1.89. The neutral gray background triggered luminance adaptation to $28$\,cd/m$^2$. 
A 2$\times$2 spatial dithering was used to avoid visible color banding in the low-contrast stimuli.
We used Python's PsychoPy toolbox~\cite{peirce2007psychopy} and a custom shader to stream frames to the display by wired HDMI connection.
All subjects were tested in a well-lit room and viewed the video display binocularly from an SR Research headrest situated 94~cm away, thus giving a field of view of $46^\circ\times20^\circ$ and a resolution of 71 pixels per degree of visual angle.
Pupil Labs Core eye trackers were mounted to the headrest to verify central gaze fixation throughout all studies.

\paragraph{Subjects}
Ten adults participated (age range 23--29, 2 female). Due to the demanding nature of our psychophysical experiment, only a few subjects were recruited, which is common for similar low-level psychophysics (see e.g.~\cite{patney_towards_2016}). All subjects in this and subsequent experiments had normal or corrected-to-normal vision, no history of visual deficiency, and no color blindness, but were not tested for peripheral-specific abnormalities. 
All subjects gave informed consent. The research protocol was approved by the Institutional Review Board at the host institution.

\begin{figure*}
  \centering
  \includegraphics[width=0.8\linewidth]{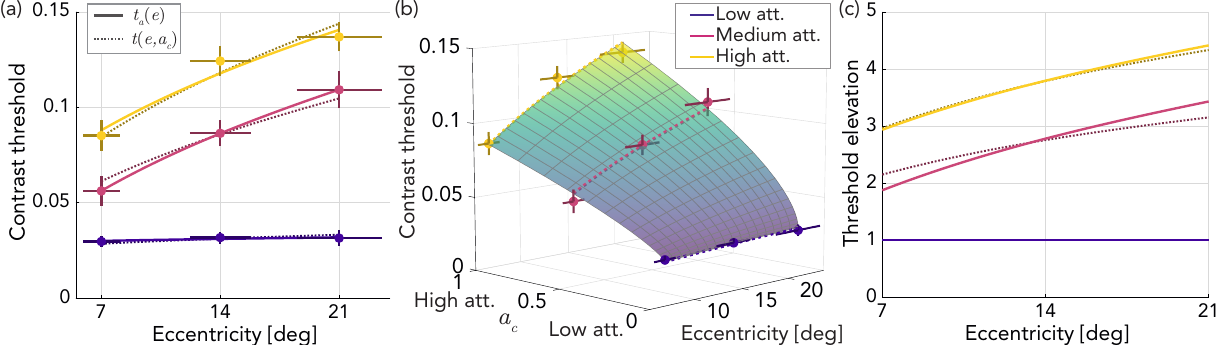}
  \caption{
  Main study:
  (a) The mean measured contrast thresholds and the fitted attention curves for the per-condition model $\thr_\load(\ecc)$ (Eq.~\ref{eq:thr}, full lines) and the unified model $\thr(\ecc,\load_c)$ (Eq.~\ref{eq:thr_3d}, dotted lines). 
  The horizontal bars display extent of the Gabors.
  The vertical error bars show standard error.  
(b) A continuous attention-eccentricity fit of the unified model $\thr(\ecc,\load_c)$ (Eq.~\ref{eq:thr_3d}).
  (c) The attention gains $\gain_\load(\ecc)$ relative to the \alow foveal attention condition computed for each of the two models.}
  \label{fig:model}
\end{figure*}

\paragraph{Procedure}
To begin the study, subjects were set up in a comfortable position on the headrest and the eye tracker was calibrated using a 5-point screen calibration~\cite{kassner2014pupil}. The thresholds for each contrast condition (stimuli No.~1--3 in Table~\ref{tab:stimuli}) were then estimated in a random order. For each condition, a two-alternative forced-choice (2AFC) adaptive staircase designed using QUEST~\cite{watson1983quest} was used to measure the contrast discrimination threshold for each attention condition, starting with the \alow, then the \amed and ending with the \ahigh foveal attention condition. 
At each step, the subject was shown a small ($1^\circ$) white fixation cross for 1.2\,s to indicate where they should fixate, followed by the attention-modulating task and contrast stimuli for 500\,ms, then a Gaussian white noise screen for 1\,s (to reduce after images).
The subject was then given 10\,s to indicate via different sets of marked buttons on a keyboard whether the target letter ``T'' was red or green, followed by whether the contrast patterns were of the same or different orientations.
If the subject failed to answer during that time, the trial would be replayed.
Each of the 3 test conditions at each of the 3 attention conditions were tested twice per subject, taking approximately 90 minutes, with subjects encouraged to take breaks between staircases.

\paragraph{Results}
Mean contrast thresholds across subjects are shown in Fig.~\ref{fig:model}a (see Supplement for table of values). 
It can be seen that the contrast thresholds are almost identical for the \alow attention condition, agreeing with the theory of \emph{cortical magnification} described by Virsu and Rovamo~\shortcite{rovamo1979estimation,virsu1979visual}.
For the \amed attention condition, however, the contrast thresholds do increase significantly with eccentricity ($p<0.05$, paired t-test between neighboring eccentricities), almost $2\times$ for $7^\circ$ and over $3\times$ for $21^\circ$ (see Fig.~\ref{fig:model}b).
Similarly, the \ahigh attention condition exhibits up to $4\times$ threshold increase within our measured eccentricity range ($p<0.05$). 
The increase in gain factors with eccentricity is consistent with work by Staugaard~et~al.~\shortcite{staugaard2016eccentricity} who showed a decrease in attentional capacity with increasing stimulus eccentricity, when stimuli are scaled in size to account for cortical magnification.
Furthermore, we observe significant differences between individual attention modes across the eccentricity ranges ($p<0.01$ for most pairs, $p<0.05$ for the \amed and \ahigh attention, paired t-test with Bonferroni correction).
This confirms our assumption that increasing the task difficulty will shift attention towards the fovea at a cost to sensitivity in the periphery.
On the other hand, despite the considerable difference between the \alow and \amed condition gradients, the gradients of the \amed and \ahigh conditions are surprisingly similar, suggesting that the effect of attention modulation is non-linear.

\subsection{Per-condition Model}

The observed increase in thresholds for larger eccentricities is nearly linear with a small distortion which we describe using the square root of eccentricity to fit attention-dependent contrast threshold models:
\begin{equation}
\label{eq:thr}
\thr_\load(\ecc) = p_0 \sqrt{\ecc} + p_1
\end{equation}
where $\load$ is denotes one of our foveal attention conditions (\alow, \amed or \ahigh).
See Table~\ref{tab:fit_params} for parameters and Fig.~\ref{fig:model}a for plots.

\begin{table}[h]
\caption{\label{tab:fit_params}Fitted parameters of our attention-aware contrast threshold model $\thr_\load(\ecc)$ (Eq.~\ref{eq:thr}). R$^2$ is the coefficient of determination.
}

\begin{center}
\begin{tabular}{cccc} 
\toprule
  $\load$ & $p_0$ & $p_1$ & R$^2$ \\
\midrule
 \alow    &  $9.672 \cdot 10^{-4}$ & \multicolumn{1}{r}{$2.741  \cdot 10^{-2}$} & $0.705$  \\ 
 \amed &  $2.737 \cdot 10^{-2}$ & \multicolumn{1}{r}{$-1.620 \cdot 10^{-2}$} & $1.000$ \\ 
 \ahigh    &  $2.714 \cdot 10^{-2}$ & \multicolumn{1}{r}{$1.612  \cdot 10^{-2}$} & $0.956$ \\ 
\bottomrule
\end{tabular}
\end{center}

\end{table}

As contrast sensitivity varies among observers, we are primarily interested in the relative attention gain represented by threshold elevations defined with respect to the \alow attention baseline condition as:
\begin{equation}
\gain_\load (\ecc) = \frac{\thr_\load(\ecc)}{\thr_{\textrm{low}}(\ecc)}
\end{equation}

Assuming orthogonality of the attention effect and other independent parameters of the stimulus, we can formulate the attention-aware contrast sensitivity as:
\begin{equation}
\label{eq:csf}
\csf_\load(\ecc,\cdots) = \csf(\ecc,\cdots) \frac{1}{\gain_\load(\ecc)}
\end{equation}
where $\csf$ is any of the CSF models discussed in Sec.~\ref{sec:lit_csf}.
In Sec.~\ref{sec:thr_validation} we use the \stela~\cite{mantiuk2022stelaCSF} model.

\subsection{Unified model}

Additionally, we explore a speculative model unifying the eccentricity $\ecc$ with a continuous interpretation of the attention condition $\load_c \in [0,1]$ where $\{\alow \to 0, \amed \to 0.5, \ahigh \to 1\}$.
We design this model as an attention-dependent sweep between the per-attention curves, parameterized relative to our lowest eccentricity of $7^\circ$. 
We model the dependency for the slope and intercept separately using two gamma curves $\load_c^{\gamma_s}$ and $\load_c^{\gamma_i}$ to account for the non-linear perception of the different attention conditions.
Due to the extreme non-linearity of the slope development we constrain $\gamma_s$ to 0.5 and fit:
\begin{equation}
\label{eq:thr_3d}
\thr(\ecc,\load_c) = \Psi\left(s_0, s_1,\load_c^{\gamma_s}\right) \cdot
\left(\sqrt{\ecc} - \sqrt{7} \right)
+ \Psi\left(i_0, i_1,\load_c^{\gamma_i}\right)
\end{equation}
to our measured data. Here, 
$\{s_0, s_1, i_0, i_1, \gamma_i\} = \{0.00243$, $0.0307$, $0.0285$, $0.0844$, $0.771\}$ 
are the fitted parameters (DoF-adjusted R$^2 = 0.973$) and $\Psi(\alpha,\beta,w) = \alpha(1-w)+\beta w$ is a linear interpolation function.

We compare the resulting unified model to our per-condition models $\thr_\load(\ecc)$ in Fig.~\ref{fig:model}a.
Despite the lower parameter count, the unified model fits the measured data within the measurement errors.
While the unified model allows for convenient interpolation, we argue for fitting task-specific models in practice, because the connection between the task and attention is highly individual and not well understood.
Hence, we use our per-condition models $\thr_\load(\ecc)$ (Eq.~\ref{eq:thr}) throughout the rest of this paper wherever not explicitly specified otherwise.

\begin{figure*}
  \centering
  \includegraphics[width=\linewidth]{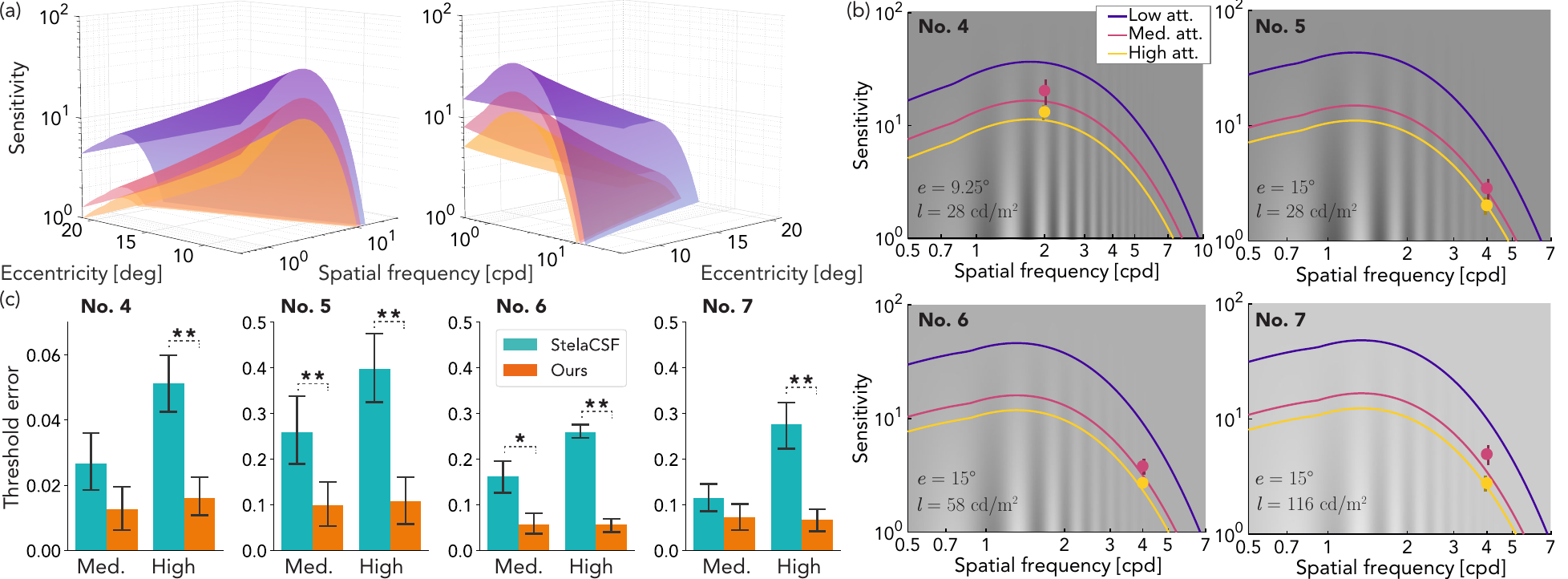}
  \caption{
 Validation studies:
(a) Two different views of an eccentricity vs. spatial frequency plot for the original \stela~\cite{mantiuk2022stelaCSF} model (the top surface, in purple) and our scaled models $\csf_{\textrm{medium}}(\ecc,\fs)$ (in magenta), $\csf_{\textrm{high}}(\ecc,\fs)$ (in yellow) for a static stimulus with an area of $1$\,deg$^2$ and an adaptation luminance of $28$\,cd/m$^2$ (same as our model study in Sec.~\ref{sec:model_user_study}).
(b) Slices of the same models describing dependency on spatial frequency for the conditions used in our validation study (see No. 4--7 in Table~\ref{tab:stimuli}).
The points denote directly measured sensitivities scaled relative to the baseline.
The bars are 95\% confidence intervals.
(c) Corresponding threshold prediction errors of \stela{} vs. our model (lower is better). The error bars are 95\% confidence intervals and significance is indicated at the $p < 0.05$ and $0.01$ levels with * and ** respectively (Wilcoxon test).
 }
  \label{fig:stelacsf}
\end{figure*}

\subsection{Validation}
\label{sec:thr_validation}

In Eq.~\ref{eq:csf}, we apply attention correction as a multiplicative factor under an assumption of orthogonality between the two functions.
If this assumption holds, the difference between new thresholds predicted by our model and their measured values should be low.

We test this by measuring the attention gains for 4 new stimuli with different cortical magnifications and adaptation luminance levels than in our main study.
We then compare the thresholds obtained by direct measurement with the thresholds predicted by our attention gain $\gain_\load(\ecc)$.

\paragraph{Experiment}
We use the same experiment procedure as for the main study, except with 2 parameter sets used to fit \stela~\shortcite{mantiuk2022stelaCSF}, a recently demonstrated unified model of CSF (see stimuli No. 4 and 5 in Table~\ref{tab:stimuli}).
These points were selected from the only 2 available datasets measured using stationary stimuli outside the fovea, with spatial frequency, eccentricity and size as different as possible to the stimuli used to fit our attention-aware CSF model.
Additionally, we test effect of varying luminance adaptation on one of these datapoints by adjusting the backlight of our display (see stimuli No. 5--7) .

\paragraph{Subjects}
Eleven adults participated (age range 23-29, 5 female), six for stimuli No. 4 and 5 and five for stimuli No. 6 and 7. Only three of these subjects participated in the main study.

\paragraph{Results}
We measured mean thresholds for the validation stimuli (No. 4--7) and the \alow attention condition as $0.032$, $0.045$, $0.57$ and $0.51$, which we use as baselines for a relative multiplicative adjustment of our measurements to corresponding predictions of \stela{}. 
We compute Interquartile Range (IQR) of this multiplicative factor to detect outliers. We treat each of the 2 per-user repetitions as a single data sample and we remove a total of 3 strong outliers with offset of 4 or more IQR from the quartiles.
We then apply this base adjustment consistently to all individual measurements to remove variability of the base sensitivity performance among users and instead focus on relative gains between attention conditions (see Fig~\ref{fig:stelacsf}b).
The resulting adjusted measurements are then compared with the contrasts predicted by the original attention-unaware \stela{} model and our derived attention-aware CSF model $\csf_\load(\ecc,\cdots)$.
We compute the error of both models in Fig.~\ref{fig:stelacsf}c.

For the 2 stimuli isoluminant with our model data (No. 4 and 5), we observe statistically significantly lower error between our and the original \stela{} predictions with respect to the experimentally measured thresholds in all conditions except for one.
The lower observed difference between \alow and \amed attention for the lower frequency stimulus (No. 4) points to an overestimation of the gain by our model here.
Notably, even in this worst case, the prediction error is still lower than that of the baseline \stela{}.

As a practical example, our measurements indicate that with \ahigh{} attention and a 28\,cd/m$^2$ display, a spatial pattern with $f_s = 2$\,cpd shown at eccentricity of $15^\circ$ will be just discriminable if rendered with an amplitude value of 32 (for a 0--255 signal range of an 8-bit display with gamma of 2.2) while our model would yield amplitude of 30 and the baseline model would adhere to the original stelaCSF prediction of amplitude 8.

The favorable performance of our model also holds for the other 2 luminance levels (stimuli No. 6 and 7). 
Despite this, we observe a trend of attention gain reduction with increasing luminance which is significant for the \amed{} attention at 116\,cd/m$^2$ ($g_\load: 3.03\to2.15$, $p<0.05$, Mann-Whitney U test) and \ahigh{} attention at 58\,cd/m$^2$ ($g_\load: 4.12\to3.37$). 
This compression could be caused by the overall increase of sensitivity under such conditions and should be considered by users of our model.

To summarize, our experiment suggests that while the assumption of full orthogonality is unlikely to hold everywhere, the relative benefit of including the attention model may still be stronger relative to the cost of this simplification.

\section{Attention-aware Foveated Rendering}
\label{sec:foveation}
The goal of foveated rendering is to reduce computational cost without introducing perceptible artifacts by exploiting the reduction of vision performance in the periphery, typically by adjusting sampling rate with respect to peripheral acuity drop (see Sec.~\ref{sec:lit_foveated}).
The quality of such foveation can be assessed by visual difference predictors, for example, \fvvdp~\cite{mantiuk2021fovvideovdp}, a state-of-the-art metric that models the spatial, temporal, and peripheral aspects of perception.
 In this section, we experimentally validate whether integration of our attention-aware perceptual model improves the performance of \fvvdp{} in predicting visibility of foveation artifacts under varying attention conditions.
To that end, we emulate a simple foveated renderer and separately calibrate the foveation intensity for three different attention regimes in a user study.
We then compare the perceptual errors of the calibrated stimuli predicted by \fvvdp{} with and without our attention-aware model to assess their agreement with human judgment.

\subsection{Measuring imperceptible foveation}

Similar to the contrast discrimination task in Sec.~\ref{sec:study}, we create a space-multiplexed comparison of foveated images.
In particular, we split the screen into left and right sides, apply foveation to one side only (randomly selected) and ask subjects which of the sides appeared more visually degraded (see Fig.~\ref{fig:foveation}a).
Then, by modifying the parameters of the foveated side, we can find the threshold for which the foveation is nearly imperceptible to the subject. 
The central transition around the fixation was replaced by a neutral background vertical bar with $6^\circ$ width and Gaussian fall-off (standard deviation of $0.5^\circ$) and the attention-modulating RSVP task, as in Sec.~\ref{sec:att_mod_task}, was then displayed centrally.

For the foveation, we base our approach on the work of Guenter et al.~\shortcite{guenter_foveated_2012} and the linear \emph{minimum angle of resolution} (MAR) model describing the reciprocal of acuity as:
\begin{equation}
\label{eq:mar}
\omega(\ecc) = m \ecc + \omega_0
\end{equation}
where the bias $\omega_0 = 1/48 ^\circ$, as in the original work, and the slope $m$ is a free variable measured as a threshold in our study.
The peripheral resolution decrease was simulated by an approximated Gaussian filter with spatially varying standard deviation:
\begin{equation}
\sigma(\ecc) = \frac{\omega / \omega_s - 1}{2 \sigma_c}
\end{equation}
where $\omega_s = 0.0283^\circ$ is the peak MAR of our display and $\sigma_c = 2$ is the chosen cut-off determining the assumed bandwidth of the filter.
Note that this particular choice primarily affects the absolute value of our slopes and not the relative ratios between conditions.

\subsection{User study}
\label{sec:fov_validation}

\paragraph{Setup}
We used the same experimental setup as in Sec.~\ref{sec:csf_study}. The stimuli consisted of one of four foveated images displayed across the entire screen with the gaze fixation directed to the center.

\paragraph{Subjects}
Thirty adults participated (age range 23--29, 10 female). Subjects were first shown the original images, to avoid exploratory saccades during the study, and then shown an example of the foveation effect.

\begin{figure*}[t]
	\centering
    \includegraphics[width=\linewidth]{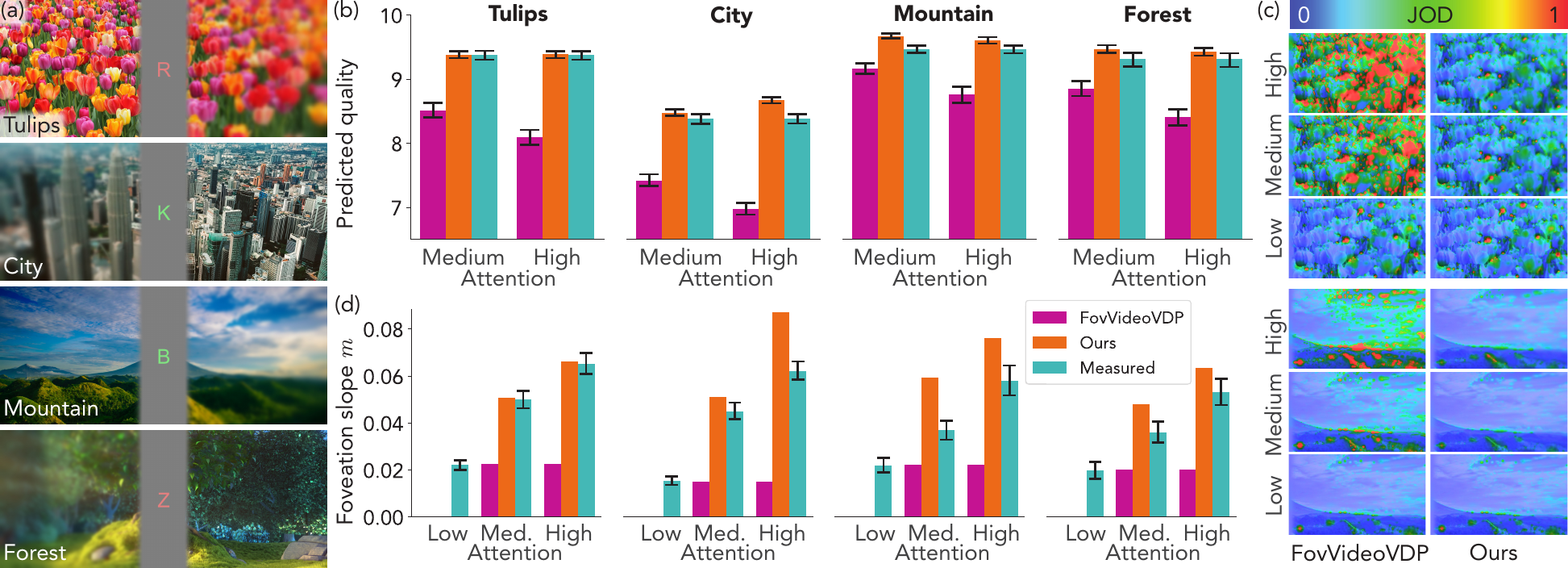}
    \caption{Foveation study: 
    (a) Stimuli from our study showing the attention-modulating RSVP task in fovea and the peripheral foveation detection task. 
    One side (randomly selected) is foveated while the other is left at full resolution.
    The foveation effect and the color and size of the RSVP task are exaggerated for visibility. 
    (b) Quality scores predicted by the original \fvvdp{} metric vs. our modified predictor (closer to \emph{Measured} is better) for the foveated images calibrated in our user study. 
    The error bars show 95\% confidence intervals.
    The \emph{Measured} quality refers to the actual quality threshold measured in the \alow condition.
    (c) Comparison of visual difference maps produced by the original \fvvdp{} metric vs. our modified predictor for the calibrated MAR slopes. 
    Colors visualize Just-Objectionable-Differences (JOD) with respect to the original ``Tulips'' and ``Mountain'' images (small section from the right periphery shown).
    (d) Comparison of MAR slopes (intensities) predicted by the original \fvvdp{} metric vs. our modified predictor for each image compared to the \emph{Measured} slopes (closer to \emph{Measured} is better).
    The legend is shared with panel (b).
    The error bars show 95\% confidence intervals of the measured values.
    Note that the model-based slope predictors do not yield variance (no error bars shown).}
    \label{fig:foveation}
\end{figure*}

\paragraph{Procedure}
As in the studies in Sec.~\ref{sec:csf_study}, subjects were instructed to always fixate on the central RSVP task and observe the foveation task concurrently in their periphery.
The thresholds for each image were estimated in a random order for each subject.
For each image, a 2-AFC adaptive staircase using QUEST~\cite{watson1983quest} was used to measure the threshold of the foveation slope $m$ for each attention condition, starting with the \alow, followed by the \amed and ending with the \ahigh foveal attention condition.
Similar to the previous studies, at each step, the subject is shown the attention-modulating task and foveation detection task for 500~ms.
The subject then indicated the color of the target letter ``T'', and if correct, were asked which side of the image (left or right) was more visually degraded.
Whenever the subject incorrectly answered the RSVP task, they were forced to start the step again.
Each subject viewed 2 images, either ``Tulips'' and ``City'' or ``Mountain'' and ``Forest'' (see Fig.~\ref{fig:foveation}a) at each of the 3 attention conditions, to keep the study duration to approximately 45\, minutes (including breaks).

\paragraph{Results}
In Fig.~\ref{fig:foveation}d, we show the measured MAR slopes $m$ averaged across the users (labeled ``\emph{Measured}'').
We applied the same IQR procedure as in Sec.~\ref{sec:thr_validation} but we did not detect any outliers.
As expected, for all images the slope significantly increases ($p < 0.001$, paired t-test with Bonferroni correction) for both \amed and \ahigh compared to \alow attention conditions.
This means that a more aggressive foveation becomes acceptable as attention shifts from periphery towards fovea. 
Furthermore, we note that there are statistically significant differences between slopes measured for at least some image pairs with \alow{} (one-way ANOVA, $F(3,116)=4.99$, $p=0.003$), \amed{} ($F(3,116)=10.26$, $p<0.001$) and \ahigh{} ($F(3,116)=3.87$, $p=0.011$) attention.
This points to a content-dependent nature of the problem.
In the next section, we discuss whether a visual difference predictor could be used to predict foveation parameters for a specific image.

\subsection{Foveated quality prediction}

\paragraph{Setup}
Acuity-driven foveation algorithms conservatively account for the worst-case scenario of the smallest detectable image detail~\cite{tariq2022noise}.
As seen in our results, distribution of contrast in specific images affects the acceptable foveation intensity.
This is modeled by visual difference predictors such as \fvvdp~\cite{mantiuk2021fovvideovdp}, which decomposes an image into spatio-temporal frequency bands and models their visibility by utilizing the CSF and a contrast masking model.
However, while explicitly modeling retinal eccentricity, the original CSF does not account for attention.
We experimentally modify the authors' implementation and integrate our model as an orthogonal scaling factor of the CSF component.
We then apply the original and the modified predictor to assess the quality of the foveated images with the per-image calibrated slopes from Sec.~\ref{sec:fov_validation}.
Furthermore, we evaluate whether an inverse process could be used to optimize the foveation intensity.

\paragraph{Quality metric}
In Fig.~\ref{fig:foveation}b, we display quality scores produced by the original \fvvdp{} metric compared to our modified predictor obtained by computing visual difference between a foveated image calibrated by each individual subject and the full-quality reference.
The scores were averaged for each image, for each of the \amed and \ahigh attention conditions.
We compare these to the expected value (labeled ``\emph{Measured}'') which was obtained by \fvvdp{} for foveated images calibrated with the \alow attention condition.
We assume that this represents the personal subject-specific threshold of the perceived quality for the given image and that it should remain constant under varying attention.

Following our previous results, we expect that images with larger objective degradation should be judged as having equivalent quality and that a successful prediction should reflect that.
Consequently, we observe that the error measured as a relative difference between our prediction and the ``\emph{Measured}'' value is consistently lower than that for \fvvdp{} ($p<0.001$, Wilcoxon test).
This suggests that our modified predictor is better aligned with the attention-modulated perception.

In Fig.~\ref{fig:foveation}c, we additionally compare maps of Just-Objectionable-Differences (JOD) produced by both predictors for the mean calibrated slopes at each condition.
\fvvdp{} indicates a strong increase of perceived artifacts even as attention towards the fovea (\ahigh attention condition).
Our method instead predicts errors on the boundary of visibility for all conditions which is consistent with our assumption.

Finally, we note that the \alow attention score predicted by \fvvdp{} based on the directly measured slopes is significantly different between at least some of the images ($9.375$ for ``Tulips'', $8.383$ for ``City'', $9.470$ for ``Mountain'' and $9.314$ for ``Forest'', $F(3,116)=150.5$, $p<0.001$, one-way ANOVA).
Since this is the baseline condition, this discrepancy is orthogonal to our primary objective of exploring the overall impact of attention, and thus we defer its investigation to future work.

\paragraph{MAR slope prediction}
\label{sec:slope_prediction}
The visual difference prediction potentially allows us to optimize foveation parameters by posing it as a constrained problem:
\begin{equation}
\Theta = \arg\min_\Theta C(\Theta)  \quad \textrm{ subject to }  \quad Q(\Theta) \geq Q_{\textrm{thr}}
\end{equation}
where $\Theta$ is a set of rendering parameters, $C(\Theta)$ is the cost of the rendering (typically time and power consumption), $Q(\Theta)$ is an image quality predictor and $Q_{\textrm{thr}}$ the required threshold.
In our case $\Theta = {m}$, $C(\Theta)$ is a monotonically decreasing function of $m$, $Q(\Theta)$ is provided by our visual difference predictor (with access to the reference image) and $Q_{\textrm{thr}}$ is obtained from the \alow foveal attention condition in Sec.~\ref{sec:fov_validation}.
The resulting problem of one variable can be efficiently solved by bisection.

Ideally, we could use a single $Q_{\textrm{thr}}$ for any image. 
However, due to the significant difference between \alow attention thresholds obtained for our images, we opt to use scene specific $Q_{\textrm{thr}}$ of $9.375$ for ``Tulips'', $8.383$ for ``City'', $9.470$ for ``Mountain'' and $9.314$ for ``Forest''.
This simulates a correction function that calibrates the underlying predictor for content-dependent effects and development of which is outside of the scope of this work.

Fig.~\ref{fig:foveation}d compares the MAR slopes obtained by solving the inverse problem with $Q(\Theta)$ implemented using the original \fvvdp{} metric and our modified predictor.
While our predicted slopes do not always match the directly measured values, for the \ahigh{} attention the errors are consistently lower than those from \fvvdp{} ($p<0.05$ for the ``Mountain'',  $p<0.01$ for the rest, non-parametric Wilcoxon test).
This is remarkable given the large domain gap between the model and foveation stimuli.
It suggests that our model is useful for attention-aware foveated rendering.
Similarly, we observe statistically lower prediction errors of our model with the \amed{} attention for the ``Tulips'' and ``City'' images ($p<0.01$) while no statistically significant differences were measured for the rest.

The remaining error could originate from a multitude of sources, among them the orthogonal assumption of CSF scaling as well as other higher level effects not accounted for in either model.
To illustrate the impact, in the worst case of the ``City'' image with \ahigh{} attention and the extreme periphery of $46^\circ$ in our experimental display setup, this error would lead to removal of spatial details in the 0.25--0.35\,cpd band which might be noticeable based on our measured data.

Importantly, we observe that the bias towards overestimation of the slopes is consistent.
This hints to feasibility of fine tuning for a specific foveation algorithm.
Even without such treatment, the relative preference of our model over the baseline is most prominent for the \ahigh{} attention which is particularly relevant for many applications where users focus at a specific target on the screen.

\subsection{Bandwidth analysis}
\label{sec:analysis}

In this section we analyze the additional computation gain that can theoretically be obtained by using our attention-aware model when the user is focused on a task in the fovea.
While we could analyze the bandwidth by decomposing the image into frequency bands and discarding the signal following the CSF predictions, we decided on a more conservative approach that instead uses our direct perceptual measurements of vision performance under the specific foveal (RSVP) task.
Therefore, we model the foveation algorithm by Guenter et al.~\shortcite{guenter_foveated_2012} together with the global mean MAR slopes $m$ obtained for the \alow, \amed and \ahigh attention conditions as $0.0198$, $0.0420$ and $0.0596$.

Unlike the discrete segmentation in the original algorithm, we simplify the analysis by assuming that sampling rate of each pixel can be controlled independently and hence we directly map the local MAR $\omega(\ecc)$ (Eq.~\ref{eq:mar}) to the computational gain $\Psi$ derived from local areal sampling density as:

\begin{equation}
\Psi(\textrm{FOV}) = \left(\int_\textrm{FOV}1\ \textrm{d}x \right) \cdot \left( \int_\textrm{FOV}\max\left(\frac{\omega(x)}{\omega_s},1\right)^{-2} \textrm{d}x \right)^{-1}
\end{equation}
where $\omega_s$ is the peak MAR of a given display with a two dimensional field of view $\textrm{FOV}$.

\begin{figure}[t]
  \centering
  \includegraphics[width=\linewidth]{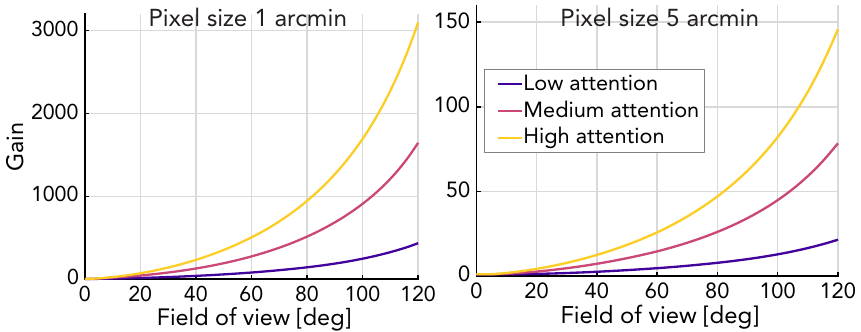}
  \caption{Computational gain analysis as a fraction of original and retained
pixel sampling density depending on pixel size and the covered visual field (horizontal axis). Note the difference
in the gain axes scales.}
  \label{fig:bandwidth}
\end{figure}

In Fig.~\ref{fig:bandwidth}, we display computational gains obtained as a function of display field of view for a common 20\,ppd (pixels per degree) and future high-density 60\,ppd displays as an upper bound for the analyzed algorithm.
It must be noted that gains in real applications are influenced by efficiency of a particular renderer.
As our contributions are independent of such design choices, more advanced foveation approaches such as noise-based enhancement~\cite{tariq2022noise} can be considered for additional gains.

\section{Discussion}
\label{sec:discussion}
The experimental data we measure and the models we fit to them further our understanding of human perception and lay the foundation of future attention-aware foveated graphics techniques. Yet, several important questions remain to be discussed.

\paragraph{Limitations and Future Work}

While our studies clearly demonstrate that modulating attention distribution between the periphery and fovea strongly impacts contrast sensitivity and foveation efficacy, we do not propose a method to measure attention.
In contrast to overt attention, which is readily measurable using eye tracking, covert attention is much more challenging.
A promising direction for exploration is the relation between pupil dilation and attentional effort~\cite{hoeks1993pupillary} and in some scenarios pupillary light response~\cite{mathot2013pupillary}.
One might also investigate the combination of eye tracking with image salience~\cite{itti2000saliency} or other metrics. 
It should also be noted that training and practice can significantly improve the ability to split attention between the fovea and periphery~\cite{zhang2022visual}.
This could lead to decrease in attention dedicated to the foveal RSVP task.
To mitigate this, we randomize trial order and encourage sufficient rest time, yet such an approach is time consuming and limits the CSF gamut that can be measured in one sitting.
While we show that our orthogonal scaling approach still leads to favorable performance when compared to baselines, we emphasize that our attention model should not be extrapolated outside of the measured eccentricities.
Finally, rather than proposing a novel foveation algorithm, we focus on demonstration of the perceptual effect as a whole.
Future work should investigate more advanced foveation algorithms and explore the effect of attention in the temporal domain.

\paragraph{Conclusion}
At the convergence of applied vision science, computer graphics, and wearable computing system design, foveated graphics techniques will play an increasingly important role in future VR/AR systems. With our work, we hope to motivate the importance of cognitive science in human perception and inspire a new axis of approaches within foveated graphics.

\begin{acks}
The project was supported by a Stanford Knight-Hennessy Fellowship and Samsung. 
The authors would also like to thank Robert Konrad, Nitish Padmanaban and Justin Gardner for helpful discussions and advice.
\end{acks}

\bibliographystyle{ACM-Reference-Format}
\bibliography{references}

\newpage
\section*{SUPPLEMENTARY MATERIAL}

In this section we provide additional discussion and results in support of the primary text.

\subsection*{Gabor patches in display space}

In Section~3.1 of the main paper we discuss the use of Gabor patches for measuring CSF. 
A Gabor patch is a complex sinusoid modulated by a Guassian envelope, defined as:

\begin{align}
	g(\mathbf{x}, \mathbf{x_0}, \theta, \sigma, f_s) = \exp \left( \frac{-\lVert \mathbf{x}-\mathbf{x_0} \rVert^2}{2\sigma^2} \right) \cos \left(2\pi f_s \mathbf{x} \cdot [\cos \theta, \sin \theta]  \right),
	\label{eq:general_gabor}
\end{align}

where $\mathbf{x}$ denotes the spatial location on the display, $\mathbf{x_0}$ is the center of the patch, $\sigma$ is the standard deviation of the Gaussian in visual degrees, and $f_{s}$ and $\theta$ are the spatial frequency in cpd and angular orientation in degrees for the sinusoidal grating function.

\begin{figure}[h!]
  \centering
  \includegraphics[width=0.5\linewidth]{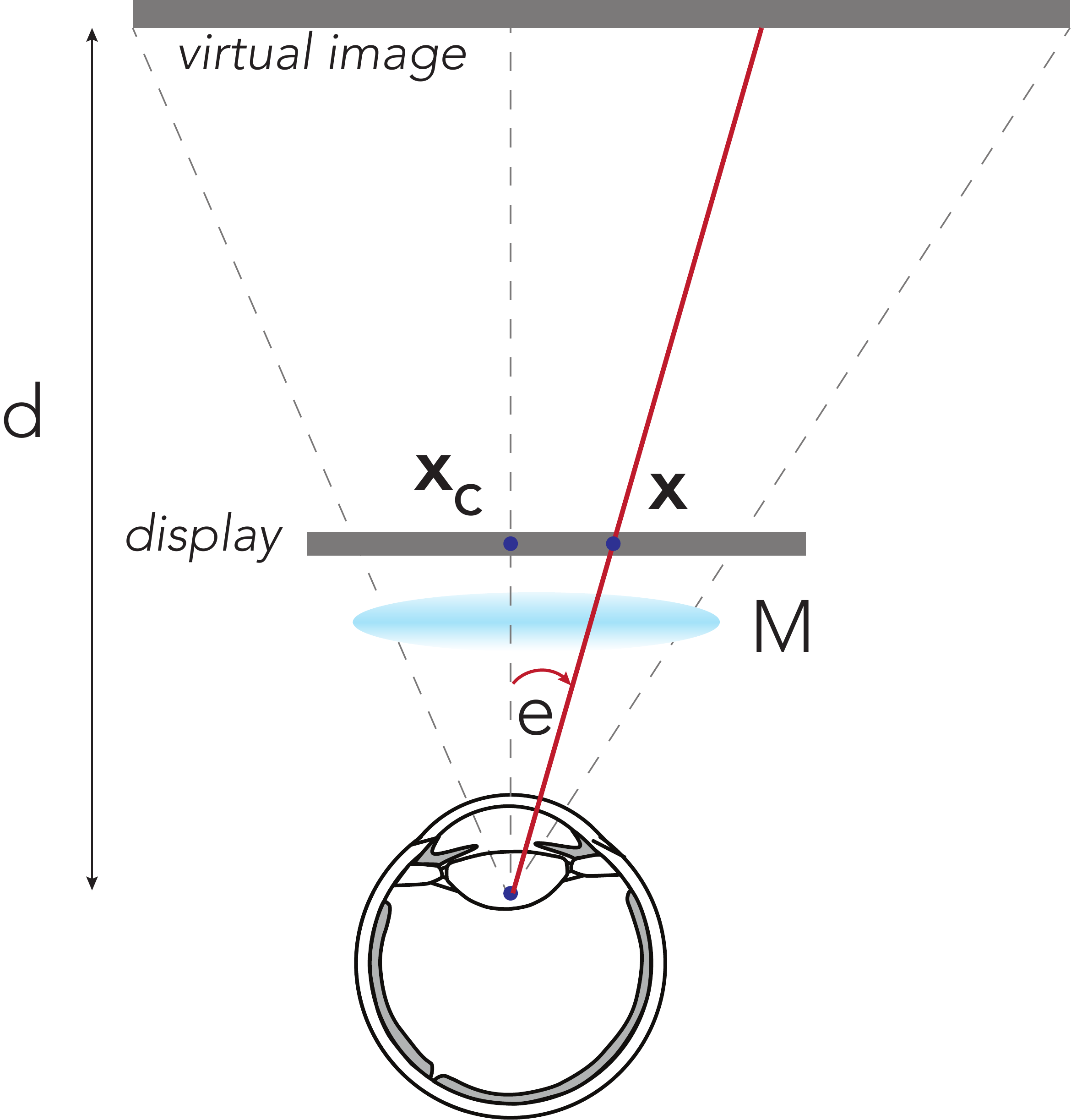}
  \caption{Schematic illustrating the geometrical relationship used to convert spatial location in pixels to eccentricity in degrees of visual angle.}
  \label{fig:px_to_deg}
\end{figure}

In our application, it is more convenient to describe spatial position in terms of eccentricity, measured in degrees of visual angle. Such transformation requires information about physical size of a pixel, $p$, and the dimensions and pixel resolution of the display (see Fig.~\ref{fig:px_to_deg}). 
Then we can derive that: 

\begin{align}
	\tan(e) = \frac{p(\mathbf{x} - \mathbf{x_C})}{d/M},
	\label{eq:px_to_deg}
\end{align}

where $e$ is eccentricity, $\mathbf{x_C}$ is the location of the pixel directly in front of the eye, $d$ is the distance to the virtual image and $M$ is the magnification factor of the lenses. 
Finally, Eq.~\ref{eq:px_to_deg} can be re-arranged and substituted into Eq.~\ref{eq:general_gabor} to re-define the Gabor function in terms of eccentricity.

\subsection*{The cortical magnification factor}

It has been argued that the changes in detection and discrimination thresholds across eccentricity can be explained by the concept of \emph{cortical magnification} \cite{rovamo1979estimation,virsu1979visual}. This model describes how many neurons in the visual cortex are responsible for processing a particular part of the visual field. The central, foveal, region is processed by many more neurons (per degree of visual angle) than the periphery. The cortical magnification is expressed in millimeters of cortical surface per degree of visual angle. We rely on the model by Dougherty~et~al.~\shortcite{dougherty2003visual} (also used by \fvvdp~\shortcite{mantiuk2021fovvideovdp}) which was fitted to fMRI measurements of V1. It is modeled as:
$$M(e) = \frac{a_0}{e + e_2} $$
where $e$ is eccentricity in visual degrees and the fitted parameters are $a_0 = 29.2$\,mm and $e_2 = 3.67^\circ$. 

Virsu~and~Rovamo~\shortcite{rovamo1979estimation, virsu1979visual} showed that the differences in detection of sinusoidal patterns and also discrimination of their orientation or direction of movement, can be compensated by increasing the size of the stimuli in the peripheral vision and the size increase is consistent with the inverse of cortical magnification. In Section~3.1 of the main paper, we follow that observation to scale our stimuli to other retinal positions such for 2~cpd and diameter of $5^\circ$ at $21^\circ$ eccentricity, the corresponding spatial frequency ($f_s$) and area ($A$) can be calculated for other eccentricity ($e$) as
$f_s = 2\cdot\frac{M(e)}{M(21)}$ and $A = 5\cdot\frac{M(21)}{M(e)}$.

\subsection*{Study results}

\begin{figure}[H]
	\centering
    \includegraphics[width=\linewidth]{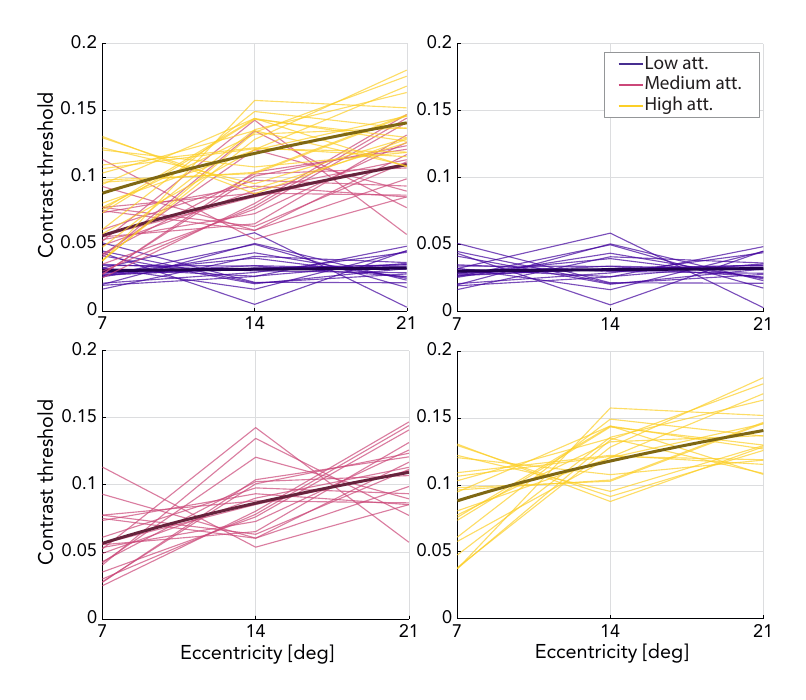}
    \caption{Contrast thresholds measured for individual subjects (thin lines) in our main study that were used to fit our model (thick lines). For clarity, the attention levels are plotted together in the first panel and separately in the other panels. 
    Mean thresholds for each plot line were rescaled to match the respective global attention level means in order to remove subject-specific variation of the base sensitivity and highlight the variation among attention levels and eccentricities. }
    \label{fig:per_user}
\end{figure}

We provide more details for the results of our experiments. 
Fig.~\ref{fig:per_user} shows trends from the main study plotted for individual users (averaged across the two repetitions). 
Tables~\ref{tbl:thresholds} and \ref{tbl:slopes} list our measured thresholds for each stimulus and measured slopes for each image.

\begin{table}[H]
\caption{\label{tbl:thresholds}Mean measured contrast thresholds for the stimuli in our main and validation studies for the different attention conditions (Sec.~3.3 and Sec.~3.5 of the main paper).}
\begin{tabular}{@{}crrr@{}}
\toprule
\multirow{2}{*}{\textbf{No.}} & \multicolumn{3}{c}{\textbf{Contrast threshold}}                                                              \\
                              & \multicolumn{1}{c}{\textbf{Low}} & \multicolumn{1}{c}{\textbf{Medium}} & \multicolumn{1}{c}{\textbf{High}} \\ \midrule
1                             & 0.0297                            & 0.0561                               & 0.0851                            \\
2                             & 0.0317                            & 0.0864                               & 0.1242                            \\
3                             & 0.0314                            & 0.1091                               & 0.1368                            \\
\midrule
4                             & 0.0325                            & 0.0607                               & 0.0905                            \\ 
5                             & 0.0452                            & 0.1304                               & 0.1806                            \\
6                             & 0.0573                            & 0.1415                               & 0.1926                            \\
7                             & 0.0508                            & 0.1059                               & 0.1832                            \\
\bottomrule
\end{tabular}
\end{table}

\begin{table}[h!]
\caption{\label{tbl:slopes}Mean measured MAR slopes from the foveation study for the different attention conditions (Sec.~4.2 of the main paper).}
\begin{tabular}{@{}llll@{}}
\toprule
\multicolumn{1}{c}{\multirow{2}{*}{\textbf{Image}}} & \multicolumn{3}{c}{\textbf{MAR slope}}                                                                       \\
\multicolumn{1}{c}{}                                & \multicolumn{1}{c}{\textbf{Low}} & \multicolumn{1}{c}{\textbf{Medium}} & \multicolumn{1}{c}{\textbf{High}} \\ \midrule
Tulips                                                & 0.0222                            & 0.0499                               & 0.0651                            \\
City                                                  & 0.0153                            & 0.0449                               & 0.0623                            \\ 
Mountain                                              & 0.0221                            & 0.0369                               & 0.0581 \\ 
Forest                                                & 0.0197                            & 0.0361                              & 0.0531 \\ 
\bottomrule
\end{tabular}
\end{table}

\end{document}